# Costs and benefits of automation for astronomical facilities

A.Yanes-Díaz[*a], S.Rueda-Teruel[a], R.Bello[a], D.Lozano-Pérez[a], M.Royo-Navarro[a], T.Civera[a], M.Domínguez-Martínez[a], N.Martínez-Olivar[a], S.Chueca[a], C.Iñiguez[a], A.Marin-Franch[b], F.Rueda-Teruel[a], G.López-Alegre[a], S.Bielsa[a], J. Muñoz-Maudos[a], H. Rueda-Asensio[a], A.Muñoz-Teruel[a], D.Garcés-Cubel[a], I.Soriano-Laguía[a], M.Almarcegui-Gracia[a], A.J.Cenarro[b], M.Moles[b], D.Cristobal-Hornillos[b], J.Varela[b], A.Ederoclite[a], H.Vázquez Ramió[b], M.C.Díaz-Martín[a], R.Iglesias-Marzoa[a], J.Castillo[a], A.López-Sainz[a], J.Hernández-Fuertes[a], D.Muniesa-Gallardo[a], A.Moreno-Signes[a], A.Hernán-Caballero[a], C.López-Sanjuan[b], A.del Pino[a], M.Akhlaghi[a], I.Pintos-Castro[a], J.Fernández-Ontiveros[a], F.Hernández-Pérez[a], S.Pyrzas[a], R.Infante-Sainz[a], T.Kuutma[a], D.Lumbreras-Calle[a], N.Maícas-Sacristán[a], J.Lamadrid-Gutierrez[a], F.López-Martínez[a], P.Galindo-Guil[a], E.Lacruz-Calderón[a], L.Valdivielso-Casas[a], M.Aguilar-Martín[a], S.Eskandarlou[a], A.Domínguez-Fernández[a], F.Arizo-Borillo[a], S.Vaquero-Valer[a], I.Muñoz-Igado[a], M.Alegre-Sánchez[a], G.Julián-CaballeroDeEspaña[a], A.Romero[a], D.Casinos-Cardo[a]

[a]Centro de Estudios de Física del Cosmos de Aragón Plaza San Juan 1, Planta 2 E-44001 Teruel Spain
[b]Centro de Estudios de Física del Cosmos de Aragón, Unidad Asociada al CSIC, Plaza San Juan 1, Planta 2 E-44001 Teruel Spain

**ABSTRACT**

The Observatorio Astrofísico de Javalambre (OAJ[†1]) in Spain is a young astronomical facility, conceived and developed from the beginning as a fully automated observatory with the main goal of optimizing the processes in the scientific and general operation of the Observatory. The OAJ has been particularly conceived for carrying out large sky surveys with two unprecedented telescopes of unusually large fields of view (FoV): the JST/T250, a 2.55m telescope of 3deg field of view, and the JAST/T80, an 83cm telescope of 2deg field of view. The most immediate objective of the two telescopes for the next years is carrying out two unique photometric surveys of several thousands square degrees, J-PAS[†2] and J-PLUS[†3], each of them with a wide range of scientific applications, like e.g. large structure cosmology and Dark Energy, galaxy evolution, supernovae, Milky Way structure, exoplanets, among many others. To do that, JST and JAST are equipped with panoramic cameras under development within the J-PAS collaboration, JPCam and T80Cam respectively, which make use of large format (~ 10k x 10k) CCDs covering the entire focal plane.

This paper describes in detail, from operations point of view, a comparison between the detailed cost of the global automation of the Observatory and the standard automation cost for astronomical facilities, in reference to the total investment and highlighting all benefits obtained from this approach and difficulties encountered.

The paper also describes the engineering development of the overall facilities and infrastructures for the fully automated observatory and a global overview of current status, pinpointing lessons learned in order to boost observatory operations performance, achieving scientific targets, maintaining quality requirements, but also minimizing operation cost and human resources.

---

[*] Axel Yanes Díaz: Further author information: (Send correspondence to ayanes@cefca.es; phone +34 978 221266)

[†1] http://oajweb.cefca.es
[†2] http://j-pas.org
[†3] http://www.j-plus.es

# 1. INTRODUCTION

There are two first main scientific projects defined to be conducted at the OAJ with the assistance of the Global Observatory Control System (GOCS[4]-[9]). They are J-PLUS[14] (Javalambre Photometric Local Universe Survey) and J-PAS[12][13] (Javalambre Physics of the Accelerating Universe Astrophysical Survey). J-PLUS[14] and J-PAS[12][13] are photometric sky surveys of several thousands of square degrees visible from Javalambre.

The Javalambre-Photometric Local Universe Survey, J-PLUS[14], is an unprecedented photometric sky survey of 8500 square degrees visible from Javalambre, using a set of 12 broad, intermediate and narrow band filters. J-PLUS[14] data releases are a powerful 3D view of the nearby Universe that is observing and characterizing tens of millions of galaxies and stars of the Milky Way halo, with a wide range of Astrophysical applications.

The Javalambre Physics of the Accelerated Universe Astrophysical Survey (J-PAS[12][13]) is a Spanish-Brazilian collaboration to conduct an innovative photometric all-sky survey of thousands of square degrees of the Northern Sky. It will observe through a set of 54 contiguous, narrow band optical filters (145˚A width each, placed $\sim$ 100˚A apart), plus two broad band filters at the blue and red sides of the optical range to reach aperture magnitude depth of AB = 22.5 — 23.5, depending on the wavelength ($5\sigma$ in $3''$ aperture). Adjacent filters have a certain overlap ensuring a spectral measurement over the whole spectrum from about 320nm to over 1050nm with 56 different spectral channels without any significant modulation as a function of the redshift. Although it has been designed and optimized to achieve 0.3% relative error photometric redshifts (photo-$z$s) for tens of millions of galaxies, enabling measurements of the cosmological baryonic acoustic oscillation (BAO) signal across a relatively wide range of cosmic epochs, the use of the narrow band filters makes J-PAS[12][13] to be equivalent to a low resolution Integral Field Unit (IFU) of the Northern Sky, hence providing the spectral energy distribution of every pixel of the sky and, ultimately, a 3D image of the Northern Sky with an obvious wealth of potential astrophysical applications. This makes J-PAS[12][13] a highly versatile project that will produce a unique legacy-value data set for a broad range of astrophysical studies of solar-system bodies, stars, galaxies, and cosmology.

Both surveys are being carried out at the Observatorio Astrofísico de Javalambre (OAJ[1][3]) using the GOCS to perform scientific operation of main process channels. On the one hand, the dedicated 2.55m Javalambre Survey Telescope (JST250), characterized by a very large Field of View (3 degree diameter), and the Javalambre Panoramic Camera (JPCam), a 1.2 Gpixel camera spanning an area of 4.3 square degrees with its 14 large format CCD mosaic and on the other hand, the Javalambre Auxiliary Survey Telescope (JAST80) is an 80cm Ritchey Chrétien telescope with a large field of view of 2 deg diameter, particularly defined for carrying out large sky photometric surveys like J-PLUS. The scientific instrument for the JAST80 telescope and for J-PLUS is T80Cam, a wide-field camera installed at the Cassegrain focus. It is equipped with a 9.2k-by-9.2k, 10μm pixel, high efficiency CCD that is read from 16 ports simultaneously, allowing read times of 12s with a typical read noise of 3.4 electrons (RMS). This full wafer CCD covers a large fraction of the telescope's field of view (FoV) with a pixel scale of 0.55"/pixel.

# 2. OBSERVATORIO ASTROFÍSICO DE JAVALAMBRE – OAJ

The Observatorio Astrofísico de Javalambre (OAJ[1][3]) is fully automated astronomical infrastructure conceived to carry out large sky surveys from the Northern hemisphere with dedicated telescopes of unusually large field-of-view (FoV). The OAJ[1][3] is located at the Pico del Buitre of the Sierra de Javalambre, in Teruel, Spain. The site is located at an altitude of 1957m above sea level, has excellent astronomical characteristics in terms of median seeing ($0.71''$ in V band, with a mode of $0.58''$), fraction of clear nights (53% totally clear, 74% with at least a 30% of the night clear) and darkness, with a typical sky surface brightness of $V \sim 22$mag $''^{-1}$ at zenit during dark nights, a feature quite exceptional in continental Europe. Full details about the site testing of the OAJ[1][3] can be found in.

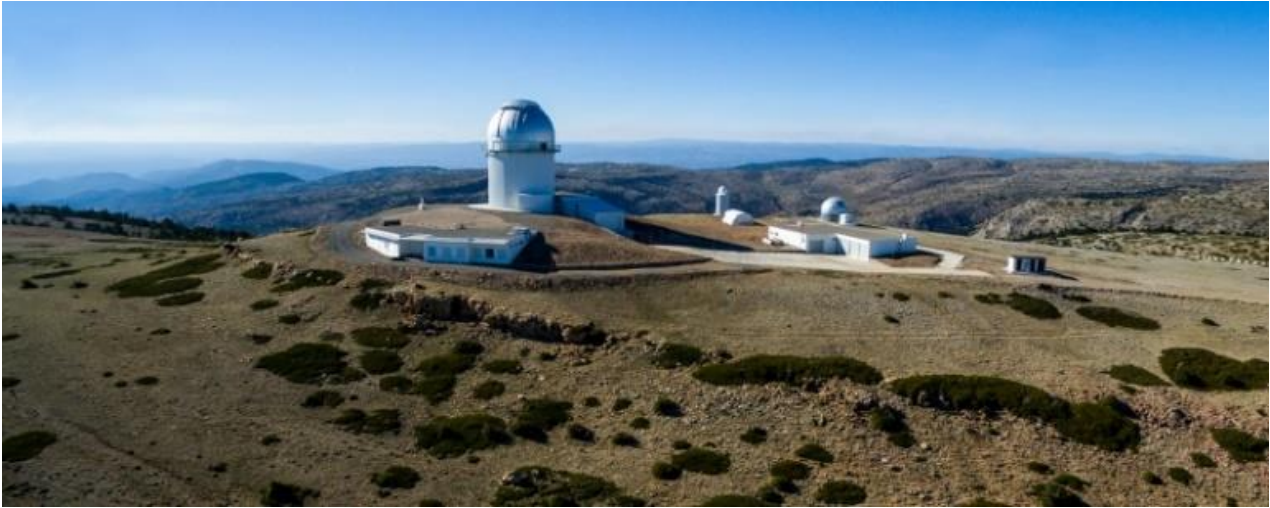

**Figure 1**

Figure 1 shows, a general view of the OAJ

The observatory is fundamentally organized around two telescopes, the Javalambre Survey Telescope (JST250), a 2.55m telescope with a FoV of 3deg diameter, and the Javalambre Auxiliary Survey Telescope (JAST80), a 0.83m telescope with a FoV of 2deg diameter. The two telescopes are equipped with panoramic instrumentation for direct imaging that allows to carry out photometric surveys with an ample spectrum of key astrophysical applications. The first defined surveys that are being carried out with the OAJ telescopes are the J-PAS and the Javalambre Photometric Local Universe Survey (J-PLUS), aimed to cover thousands of square degrees of the sky visible from Javalambre with an specific set of 56 narrow-band, contiguous, optical filters (J-PAS) and 12 broad-, medium-, narrow-band optical filters (J-PLUS).

Automation and control engineering play a key role in boosting observatory operations. For this reason, the OAJ[1][3] was planned from scratch to be a fully automated astronomical observatory in order to perform robotic operations for the whole observatory processes, of course including science operations on the top of the global observatory process. The general idea is to perform efficient operations from a global point of view for the whole observatory processes, maximizing observatory operations performance, achieving scientific objectives, maintaining quality requirements but also minimizing resources, materials and human interaction.

Apart from the telescopes, scientific instrumentation, GOCS[4]-[9] and general infrastructure, the OAJ[1][3] also includes the data center *Unidad de Procesado y Archivo de Datos*[15] (UPAD) with capacity to provide reduced and calibrated data on a quasi-real time basis and to archive and allow external access to the whole scientific community to images, data and UPAD generated catalogues.

The definition, design, construction, exploitation and management of the observatory and the data produced at the OAJ[1][3] are responsibility of the Centro de Estudios de Física del Cosmos de Aragón (CEFCA). The OAJ[1][3] project started in March 2010, mostly funded by the *Fondo de Inversiones de Teruel*, a programme supported by the local Government of Aragón and the Government of Spain, and is essentially completed since 2015. In October 2014, the OAJ was awarded with the recognition of Spanish ICTS (*Infraestructura Científico Técnica Singular*) by the Spanish Ministry of Economy and Competitiveness.

# 3. TWO DIFFERENT APROACHES FOR CONTROL SYSTEM

There are many ways to solve the challenging problem of designing and implementing a high-performance astrophysical observatory from scratch, but also, there are several aspects to consider when it comes to design a new scientific infrastructure. In this paper we are going to focus on the key role of automation and the ways to achieve our goals through it.

When it comes to engineering automation design of the observatory, there are two approaches for control system design: The classical splitted way of "**OCS et al**" and the comprehensive way of "**GOCS**".

On the one hand, the **classical approach** is the splitted way of "**OCS et al**". The OCS, that stands for Observatory Control System, which is mainly a software to perform observations focused on controlling scientific systems, such as scientific instrument and its telescope, in order to perform scientific operation through scheduled observations and processing sequential data management. In this case, the word observatory is confusing because it only refers to telescope and camera systems, exceptionally including any extra system, such as dome or meteorological station. All the rest of observatory systems are controlled from a substantial amount of different independent interfaces, not interconnected among them and completely separated from the OCS.

On the other hand, there is a **comprehensive approach** for control system design as an all-inclusive concept for all and every single system deployed in the observatory "**GOCS**". The GOCS[4]-[9] stands for Global Observatory Control System, it is an integrated software-hardware infrastructure that integrates all systems located at the observatory. GOCS uses a set of rules and regulations grouped as a model that unifies fundamental characteristics of systems, procedures and control flows to provide a single tool that simplifies global operation of the observatory. It implies a bigger effort to implement, not only to set up GOCS infrastructure, but also due to the fact that normalization is integrated at system level, however, benefits far outweigh this effect.

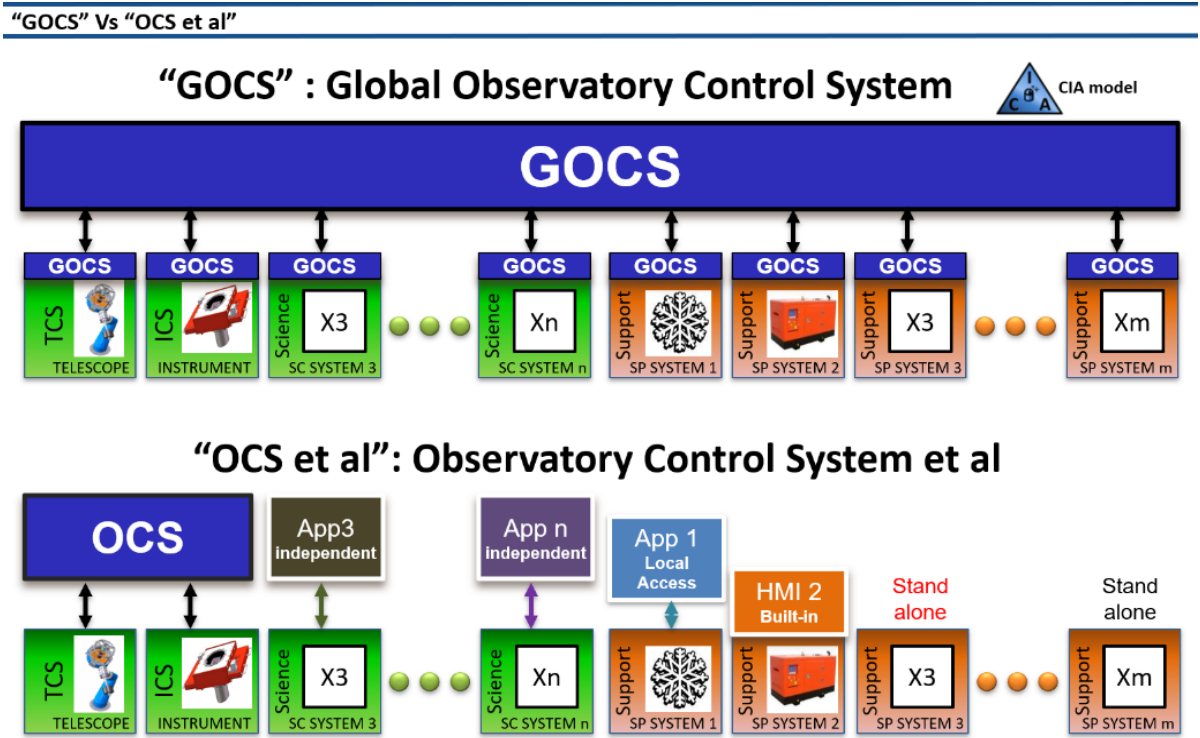

**Figure 2**

Figure 2 shows, a general overview of these two different approaches

# 4. GOCS

The GOCS[4]-[9] idea was developed by CEFCA in 2011 using CIA[4]-[9] model in order to set up the OAJ as a fully automated astrophysical observatory.

The relationship between all systems deployed at the observatory has to be optimized in order to facilitate coordination and best performance for observatory global functionality. For this reason, CIA model gives a set of rules and regulations, to use in the process of designing and implementing a full automated astronomical observatory, considering it as a whole and focusing on overall efficiency. It is basically integrated by several systems grouped into entities, coordinating all systems to achieve common objectives of science or support. These entities are known as "Process Channels". On the one hand, Science Process Channel are composed by a coordinated group of systems in charge of producing scientific data (just a few of them represented in green in Figure 3). On the other hand, Support Process Channel are integrated by a coordinated group of systems in charge of supplying basic observatory structural services to support the whole set of observatory needs (just a few of them represented in brown in Figure 3).

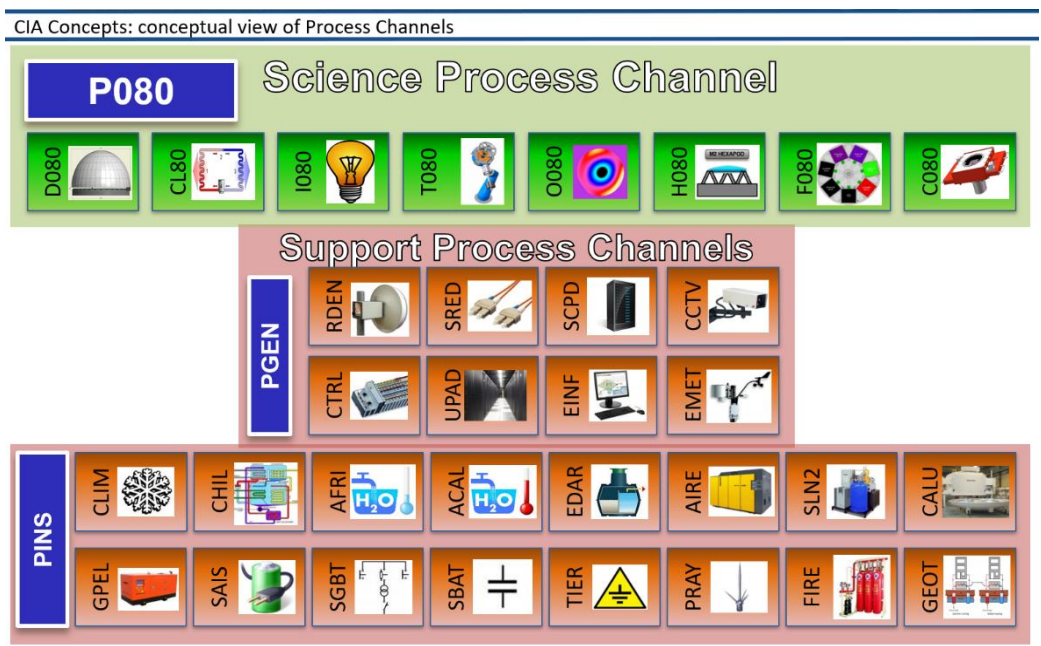

**Figure 3**

Figure 3 shows, a small representative sample of observatory`s systems grouped together to be coordinated into different process channels: P080, PGEN and PINS

CIA[4]-[9] model summarizes the set of minimum requirements to fulfill the design of the GOCS[4]-[9] in order to add and obtain high performance in the observatory's functionality. As we have seen, it extends the functionality of classical OCS to achieve the main goals by adding the specific requirements. In the CIA[4]-[9] concept all systems at the observatory must be included as a global idea, not only those related directly with astronomy are important, but also those related with general infrastructure. This is our main premise because all systems are interrelated and as a matter of fact, they have a real dependency in the deployment of the operational model of the observatory. CIA[4]-[9] also includes the goal of giving added value and functionality to all staff profiles working at the observatory and covering at the same time these services: Operations (Scientific, Support, Service, Technical, Maintenance), Management, Key Performance Indicators (KPI), Engineering and Exploitation.

These are the five types of operation that we take into account in GOCS[4]-[9] to address the complete operation of the observatory:

- **Scientific operation**: Performing activities using **scientific systems** through the provided operation interfaces HMI in order **to obtain scientific data.**
- **Support operation:** Performing activities using **support systems** through the provided operation interfaces HMI in order **to supply and support scientific/support operation itself**.
- **Service operation:** Performing activities using **any kind of system** through the provided operation interfaces HMI in order **to obtain engineering data for system improvement or system check.**
- **Technical operation:** Performing specific actions and procedures on **any kind of system** through specific procedures and interfaces only for engineers and technicians, depending on the system, in order **to upgrade, update, change over or modify the system or part of it.**
- **Maintenance operation:** Performing specific actions and procedures on **any kind of system** through specific procedures and interfaces only for engineers and technicians, depending on the system in order **to upkeep the system working with full functionality, availability, high performance and/or quality.**

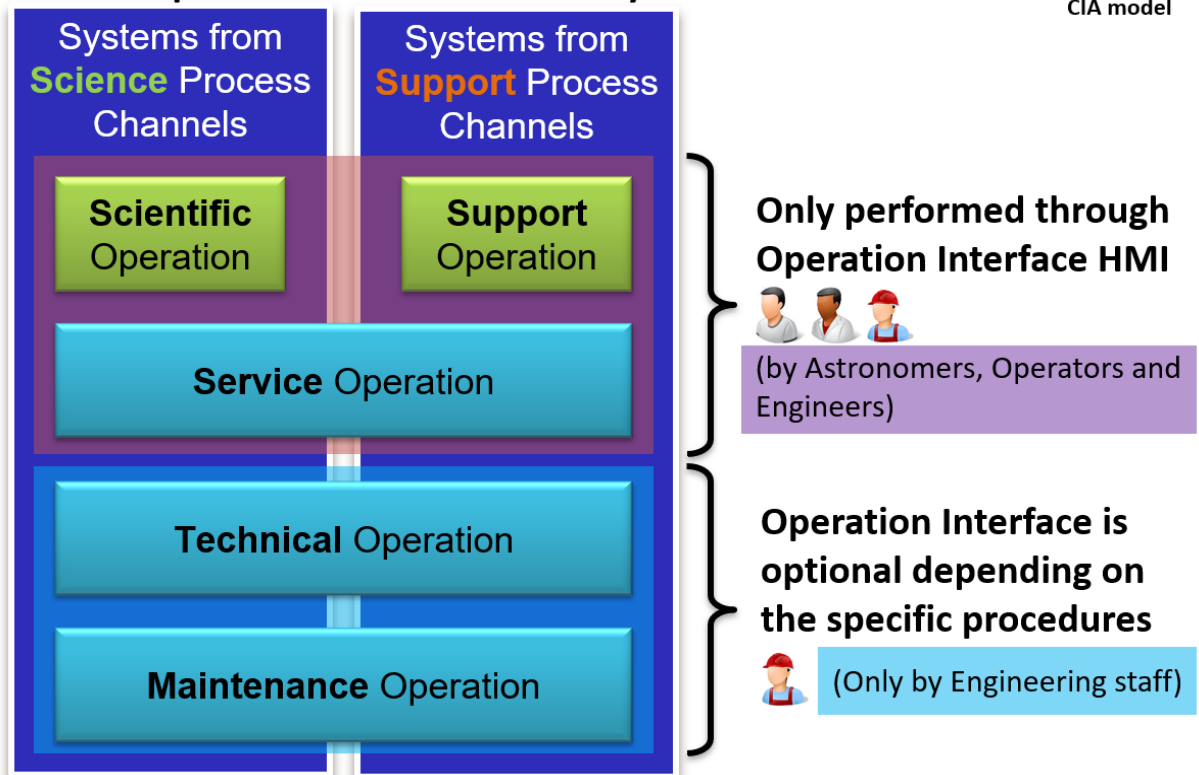

**Figure 4**

Figure 4 shows the five types of operation in GOCS to address the complete operation of the observatory

## 4.1 GOCS and Control Integrated Architecture

**Integration Layers:** The main concept is to integrate of GOCS[4]-[9] all levels implementation to perform full operation of the observatory and controlling its systems from low-level (physical level) to high-level (application level). This is a hardest concept of hardware/network/software flexibility because an observatory is an alive system that it is changing every day as well as in different project periods, so preparing a good infrastructure in advance in order to adapt easily to the most demanding changes is vital. If we consider these layers as horizontal layers, at any time we have to be able to vertically include or remove features and functionalities in the global system. This is to say that the hardware layer is "almost a software layer" because the basic infrastructure, rules and regulations are defined from scratch to obtain flexibility for quick adapting to new needs.

- **Software layer:** High-level programmable layer for controlling all systems at the observatory.
- **Network layer:** Communications layer interconnecting systems with hardware layer and software layer.
- **Hardware layer:** Low-level programmable layer for controlling each system at the observatory.

GOCS has been designed with EtherCAT[17] technology, a deterministic fieldbus network connecting all of the buildings at the observatory via multi-mode optic glass fiber 50/125 µm ring. EtherCAT[17] enables control concepts that could not be realized with classic fieldbus systems. The high bandwidth enables status information to be transferred with each data item. The bus system is not the "bottleneck" of the control concept. Distributed I/Os are recorded faster than is possible with most local I/O interfaces. A second fiber optic network is also installed at the OAJ in this case with redundant star topology for Ethernet communication between buildings. In both cases the selected fiber optic immunizes the network layer infrastructure against occasional rays.

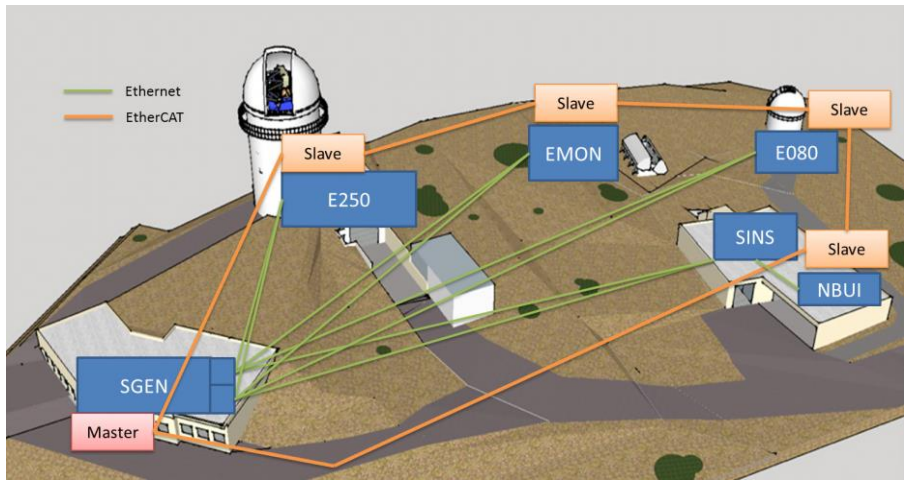

**Figure 5**

Figure 5 shows a representation of the main EtherCAT and Ethernet networks and nodes; secondary networks are not represented in order to facilitate interpretation.

On each building there is a Beckhoff[16] EtherCAT slave node connected to the main EtherCAT[17] network. On the one hand this node is communicating data with a secondary EtherCAT network interchanging all signals coming from all different PLC systems present in each building, on the other hand this node is communicating data with the main Beckhoff[17] PLC which is the master of the main EtherCAT network and is in charge of concentrating all signals at the observatory in order to interchange data with SCADA.

At the observatory there are a wide range of very different systems and subsystems. Figure 6 shows the heterogeneity in the classification of all systems types installed at the observatory. There is also a scheme explaining GOCS modules and how they are interconnected. This global scheme represents the GOCS technical structure of the CIA model implementation.

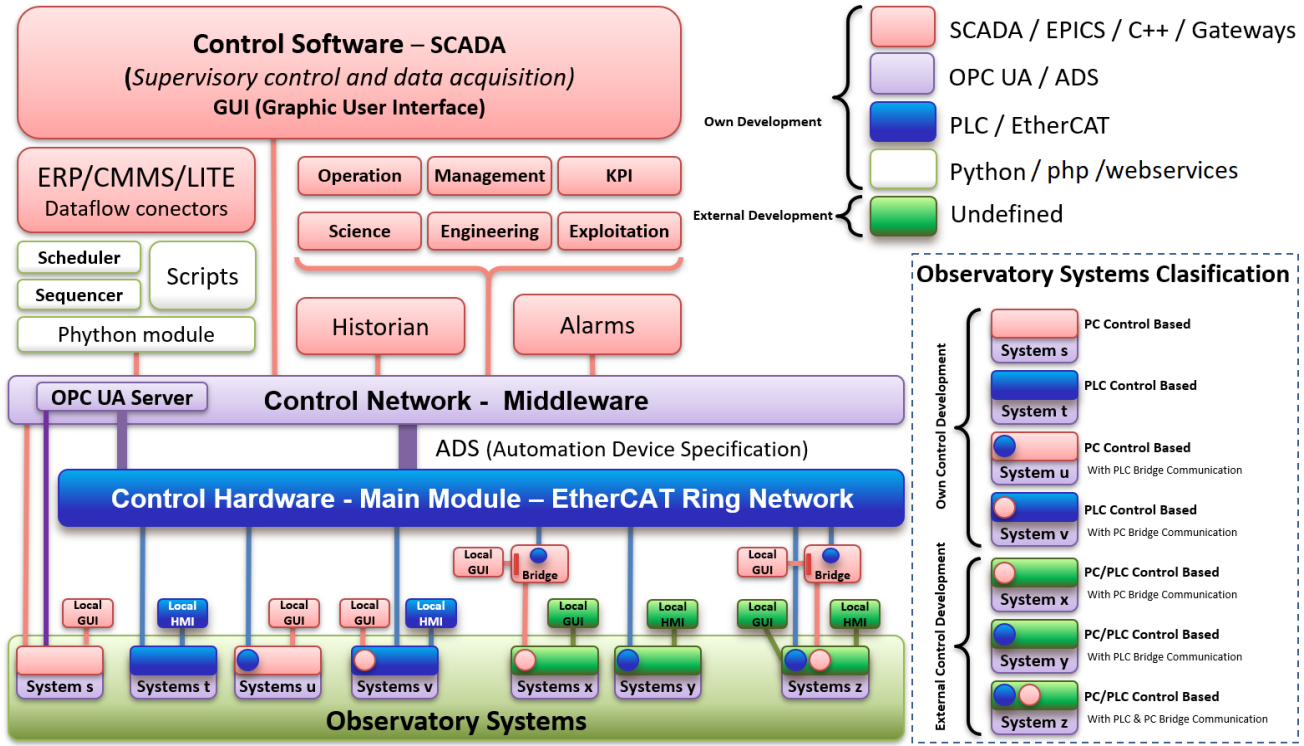

**Figure 6**

Figure 6 shows GOCS modules implementation and the seven types of heterogeneous observatory systems

## 5. DIFICULTIES

In any type of control engineering project, it is quite common to find many on a daily basis and long-term usual difficulties with normal relative importance that have to be solved on time as soon as the team find and tackle them. But in this article, we are going to mention two issues that have had important relevance for us. One is related to HR planification and the other one related to technical aspects, but in the end affecting to HR planification as well.

The main difficulty we have encountered is related to HR planning and availability during the different stages of the project.

On the one hand, at the level of administrative resource planning, given that, when you start up an astrophysical observatory from scratch, the amount of administrative and bureaucratic work that you have to do is extremely high.

On the other hand, the initial approach of a very small workforce greatly limits the margin of action when appear certain unexpected events, having to strictly distribute the resources available at all times by priority.

Later in another section of this article, we will make a more detailed mention of the human resources that we have had available during different stages of the project. Not only is there important work at the engineering level, but also at the administrative level there are a substantial number of human resources that are required, which in our case we have absorbed this work directly with the personnel of the engineering department.

Human resources must be carefully planned and evaluated periodically during the most important phases of project development and implementation until regular operation of the Observatory is achieved.

Another important difficulty that we encountered during the phases of implementation and integration of the observatory's control systems was that some of the Observatory's systems were black boxes. In the sense that they were completely inaccessible proprietary systems in which we had serious technical difficulties in directly integrating the control system. Luckily, these types of systems have been few, because we already take care that the requirements of external contracts always comply with the CIA integration rules. However, for various reasons, sometimes it is not always possible to meet these needs, so in some of these cases we solved them with parallel control engineering and even reverse engineering. Finally, we were able to solve all the integrations but it required some additional effort. For this reason, we must plan control engineering resources with sufficient margin, at least 6% overtime, to take this type of contingency into account.

## 6. BENEFITS

All "GOCS" approach benefits could be resumed in a simple sentence: "GOCS minimizes Errors, Resources, Time and Costs while maximizing observatory operations performance, achieving scientific objectives, maintaining high quality requirements". These perks are interconnected and they based on the natural convergence of CIA model benefits adding efficiency to the traditional approach.

**GOCS reduces errors** because in the end it is reducing the number of human interactions in the platform, but also reducing the number of interfaces, tools and apps to operate. GOCS is not only affecting to departments directly related with observatory operation, also shares data among other departments making easier interdepartmental processes in the organization.

From operation point of view, we are reducing the number of human interactions because basically users are operating just a single system, which is called "Observatory". In the end, with the minimum number of possible transactions and interactions in only just one tool and one interface. Therefore, reducing error sources.

Of course, the observatory is a complex system, but deep down, if we analyze it well, it is reduced to a ramified grouping of new systems in which they are becoming a little simpler each time. On each grouping level we use the common characteristics that serve us to group the processes of the operation and so on. This fact gives us a new advantage, the capability of having a very simple high-level interface for usual daily working mode "normal process" (NP) in science operations minimizing human errors but, at the same time, maximizes tool power having a low-level interface, with deep detailed functionality for engineering operations.

**GOCS allows material resources, time and costs to be minimized** because the complete and detailed operation of the entire observatory is recorded in the tool. And this is a powerful advantage, for example, to link processes that are apparently independent. With very little effort, the engineering team can optimize observatory processes in order to reduce energy or fungible resources. They can analyze any process to improve its efficiency and relate it to any type of event, process or circumstance that occurs in the observatory and at first seems to have nothing to do with it.

**GOCS reduces required staff for observatory operations** due to it optimizes processes in an evolutionary way using quick reengineering capabilities, freeing up a lot of necessary functions that are automated in the operation process and therefore the need for human resources decreases to the extent that all the necessary processes of the observatory can be carried out with very small permanent workforce. The simple fact of having the whole set of observatory's real time data and the historic values under the same tool gives a powerful capacity of analysis to standardize and normalize observatory operations.

**GOCS reduces time to operation** permitting a gradual operation start. Once we have implemented the basic infrastructure of GOCS, we can make a quick implementation to start doing the global operation before having the full functionality of the tool. This is a powerful advantage because it greatly facilitates the minimum time to start operating a certain process channel once we have been defined the minimum requirements.

**GOCS reduces time and cost** optimizing the performance of operations in the observatory, because it allows the integration of the data flow at two levels, combining long-term data from planning with instant data such as the execution of operations in real time. A clear example of this would be the coordinated management of the observatory's general inventory, where interdepartmental long-term and real time needs are clear and evident. Another example would be the management of direct purchases. For example, supplies whose levels are measured and updated in real time in the GOCS records and can generate automatically a purchase order. These aspects have a direct impact on downtimes and maintenance management of the facilities and simultaneously on the administrative and accounting aspects. These examples are clear evidences of direct improving operations performance.

## 7. CURRENT STATUS

At the time of writing this paper, the development status of the OAJ GOCS is 92% completion. Currently we are performing the global operation of OAJ[1]-[3] but there are still pending to implement some part of the 3 top level functionality.

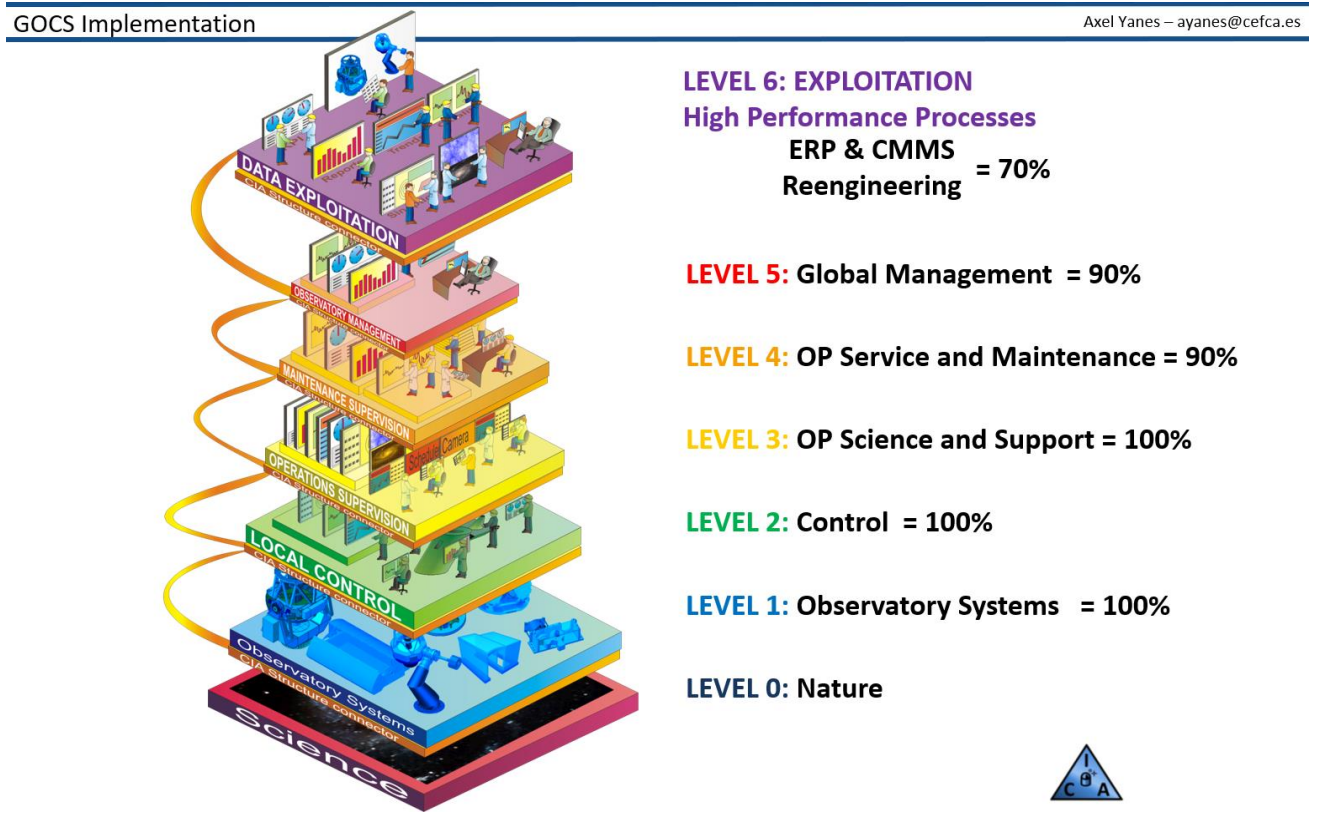

**Figure 7**

Figure 7 shows GOCS levels of implementation representing different layers of functionality

Below, we show in a summarized way the current status of all the implementations developed and commissioned in this project.

The development and implementation of OAJ control system has 6 different overlapping levels: Level1: Observatory construction and observatory systems, Level2: Independent and Local control systems, Level3: Integrated Architecture for NP (Normal Process working mode to perform Science and Support operation), Level4: Global Process Channels for NP+SM (added to Normal Process working mode there is available also Service and Maintenance working mode), Level5: Global Observatory Management, Level6: Exploitation: High performance coordination in order to interchange planification and logistic data within the organization. Levels 1, 2 and 3 are finished and Levels 4, 5 and 6 are still pending to be finished, mainly depending on Enterprise Resource Management integration and computerized maintenance management system integration, but with expectative of being completely terminated in 2024. Currently we are working on POAJ and P250 process channels. The detailed percentage of completion are the following:

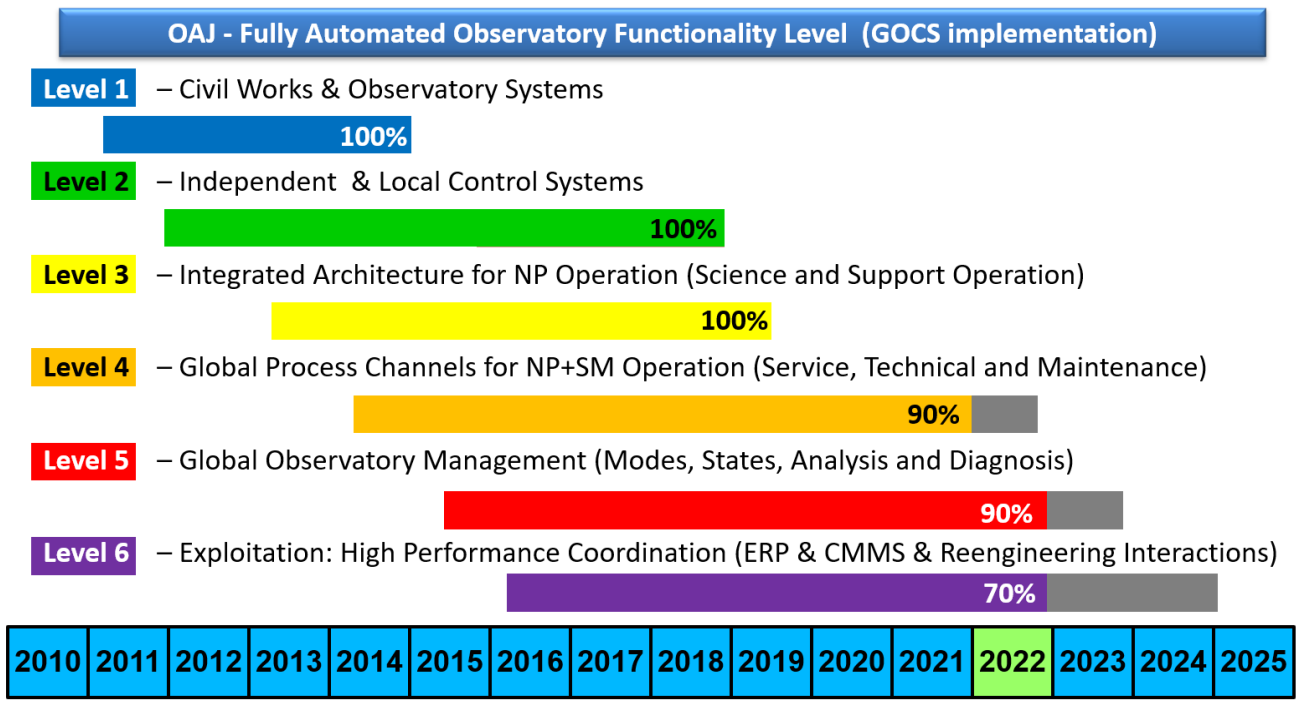

**Figure 8**

Figure 8 shows Fully Automated Observatory Planning and current status

Regarding support process channels, we are in a continuous period of reengineering because the systems are integrated but they even require some process optimization, basically, now we are mainly focused in secure some of our supplies like liquid nitrogen supply and glycol water system by mean of redundant systems.

From a development status point of view, the Support Process Channels for P250 are already deployed and they are working properly in a partially coordinated mode, so there are still pending some extra developments in the control system to achieve full automatic operation in order to support robust operation of the whole set of Science Process Channels, such as P250.

Considering the science process channels. On the one hand, P080 is fully operational achieving a very good performance, basically we are now doing small reengineering tasks and above all we are focus on analyze its strengths and weak points in order to define an improved version of this process channel. On the other hand, P250 has been recently updated to work with JPCam and currently finished technical commissioning and starting scientific operation finishing its second milestone with JPCam instrument. There is still pending future steps to integrate and perform fully automatic operation of P250 and POAJ as global operation for the observatory.

Currently we are working hard to implement High Performance Coordination functionality that will be carried out through the integration of global technological platforms: On the one hand there is the ERP that stands for Enterprise resource planning, which is the integrated management of core business processes, often in real-time and mediated by software and technology, that facilitate the coordination of the economic administrative function of CEFCA with the operational needs of the observatory. On the other hand, there is the CMMS that stands for Computerized Maintenance Management System, which provides the technology platform for management of maintenance resources for the critical subsystems of the OAJ[1]-[3]. As a result, the goal is to perform as soon as possible the efficient global operation of observatory process channels within a global framework of intercommunication with administrative processes.

# 8. AUTOMATION COSTS FOR OBSERVATORIES

Following, we are going to study in detail what the breakdown of the total cost of the GOCS control system would be. To have a better estimation of this investment, we are going to review the details in two detached sections. On the one hand, in economics terms for supplies costs and on the other hand, the manpower necessary to carry out the development, implementation and start-up of the complete system with full functionality.

The results shown in this section are obtained taking into account our own investment made in the OAJ. Take into account that represented expenses had been invested during an extended period of time while the observatory civil works was starting (2010), until acquiring the full set of observatory systems (2018). The equivalent accumulated inflation for this period of time updated to 2022 in Spain is 23.4%. For 2022 investments estimations you should apply this percentual increment. Therefore, a supposed total OAJ investment in 2022 would be 39.5M€ instead of 32M€.

Following, we are going to analyze the total cost of automation investment for the Observatory compared to the total investment of the observatory itself and the breakdown of automation expenditures.

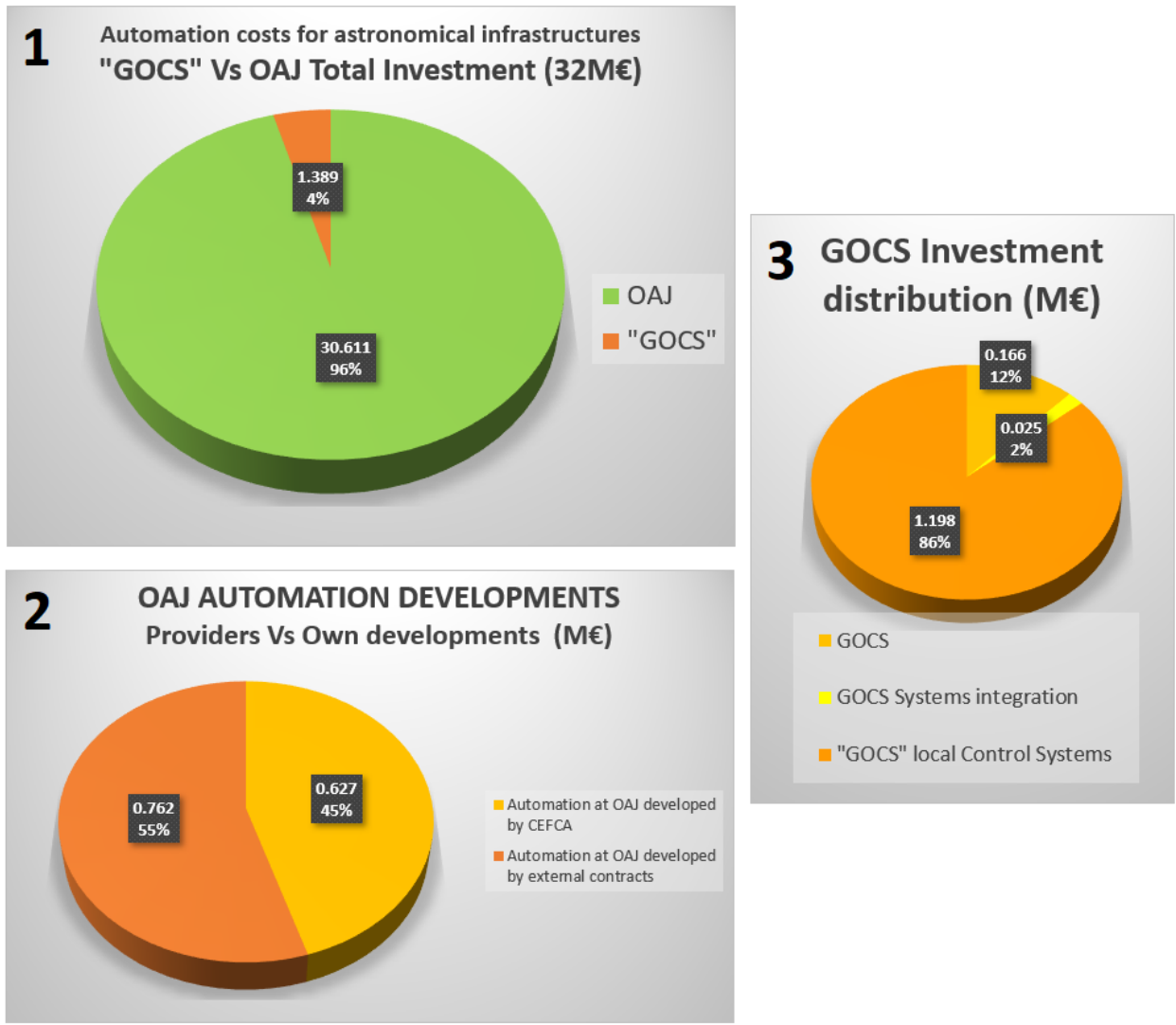

**Figure 9**

Figure 9 shows the breakdown of automation cost for OAJ

Figure 9.1 The cost of automation represents less than 5% of the total investment of the observatory, which includes not only GOCS itself, but also, absolutely everything deployed in the observatory in terms of control systems engineering. Ranging from very simple systems, even just part of it, up to the most complex systems in charge of controlling a process by automatic means.

Figure 9.2 The pie chart represents the distribution between direct investment made by CEFCA for own control system engineering and the subcontracted ones. As a matter of fact, 55% represents a considerable set of supplied observatory systems, which are given intrinsically with own control systems. The key point for these supplied systems is to have an excellent technical requirements description at the request for quotation process of those systems, following CIA model rules and regulations to facilitate, as much as possible, later integration in GOCS. In order to have a more realistic estimation of project costs, take into account that even in these subcontracted projects, there are a substantial amount of work needed to be done by your own control staff. This fact will be remarkable later when we study control engineering manpower.

Figure 9.3 The pie chart represents the GOCS investment distribution. It outstands that the main contribution, by far, is due to all systems deployed at the observatory. Their local control systems embodies an 86% of the total "GOCS" approach investment.

Following, we are going to analyze the differences of these two approaches "GOCS" Vs "OCS et al".

Figure 10.1 This figure highlights that both investment approaches are quite similar in terms of economical budget due to the fact that the dominant factor setting the investment budget is the set of control systems deployed at the observatory, as we could see previously. Therefore, there is only a 3% of difference between both approaches.

Figure 10.2 The pie chart highlights the main difference in both approaches, which is directly the tool. In this case the dominant factor setting the investment budget is the GOCS infrastructure including hardware, network and software layers. Basic infrastructure is splitted into groups depending on its localization inside the Observatory. This is the reason why exists five mains Control-Nodes (CN). They are the following: Installations Building (BINS), T080 Building (B080), Monitors Building (BMON), T250 Building (B250) and General Building (BGEN). All these nodes are linked thanks to an EtherCAT Ring topology network and Ethernet Star topology network as we can see in Figure 5. This strategy gives a stronger redundancy and error identification thank to bi-directional communication channel. These Communications are implemented by Multi-Mode Optic Fiber, using a protocol and technologies from EtherCAT and Ethernet. In both cases there are different control branches that are implemented by copper cable spreading out from each node. Fibers, from each node, are centralized on the Main Server Rack inside the General Building. The total distance that the fiber goes through the network is about 350 meters long. The network topology is configured on the Main Server Rack. The Central PLC is located at the General Service Building. All systems, including acquisition and actuator devices are controlled using different EtherCAT Bus Extensions through each node. Apart from these elements of the infrastructure there are hardware for local and remote control rooms, for Inductive Automation Ignition[18] SCADA and redundant OPC UA servers.

Figure 10.3 The last pie chart breakdown both approaches and having analyzed the first dominant factor (the tool). We focus on systems integration as secondary factor, putting emphasis on the fact that integration for "OCS et al" is almost negligible compared to "GOCS" approach. This is due to the fact that "OCS et al" integration is mainly concerning to telescope and instrument while "GOCS" integrates all systems.

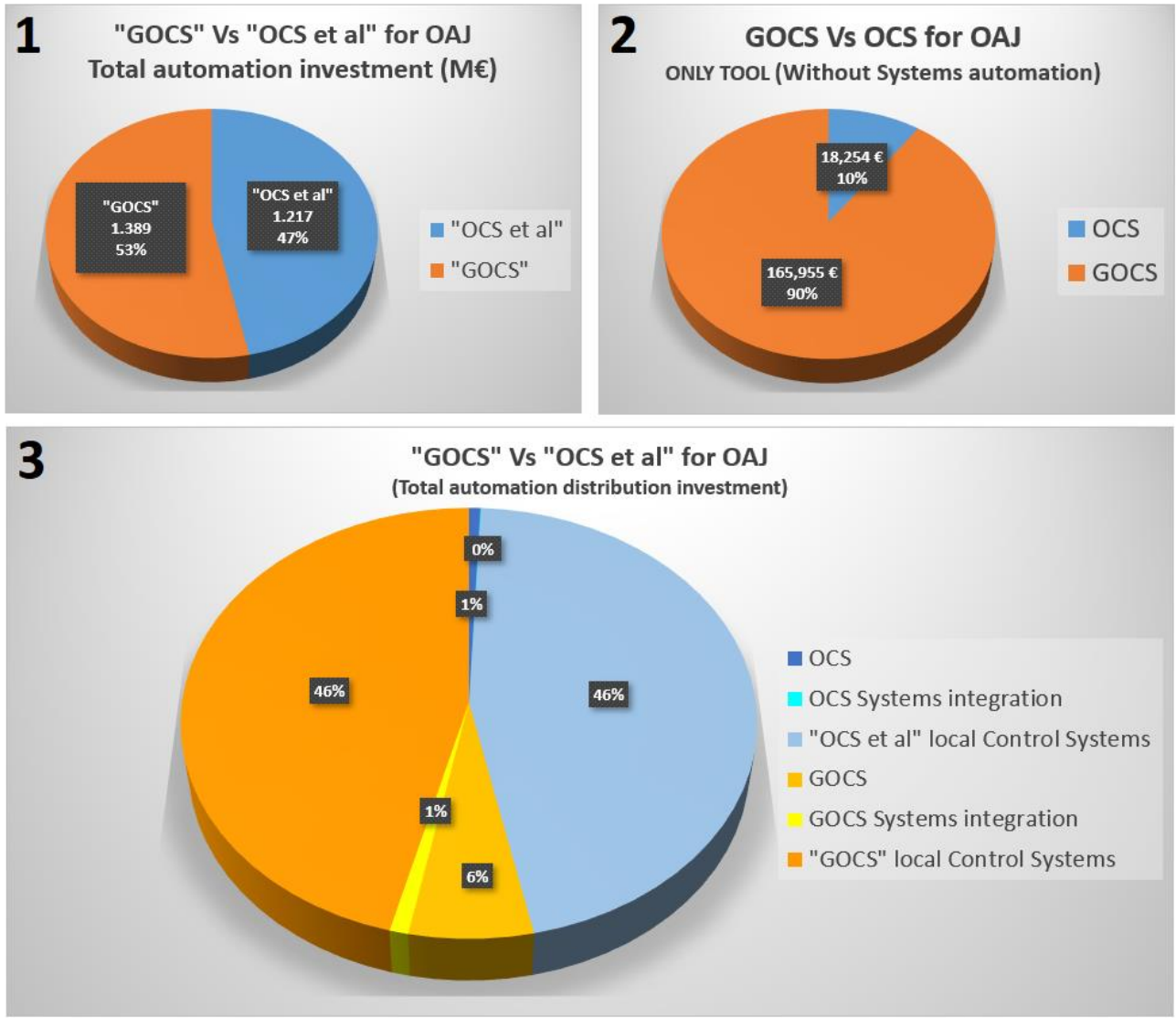

**Figure 10**

Figure 10 shows three pie charts breaking down automation investments for astronomical infrastructures comparing "GOCS" and "OCS et al" for a better understanding of both approaches

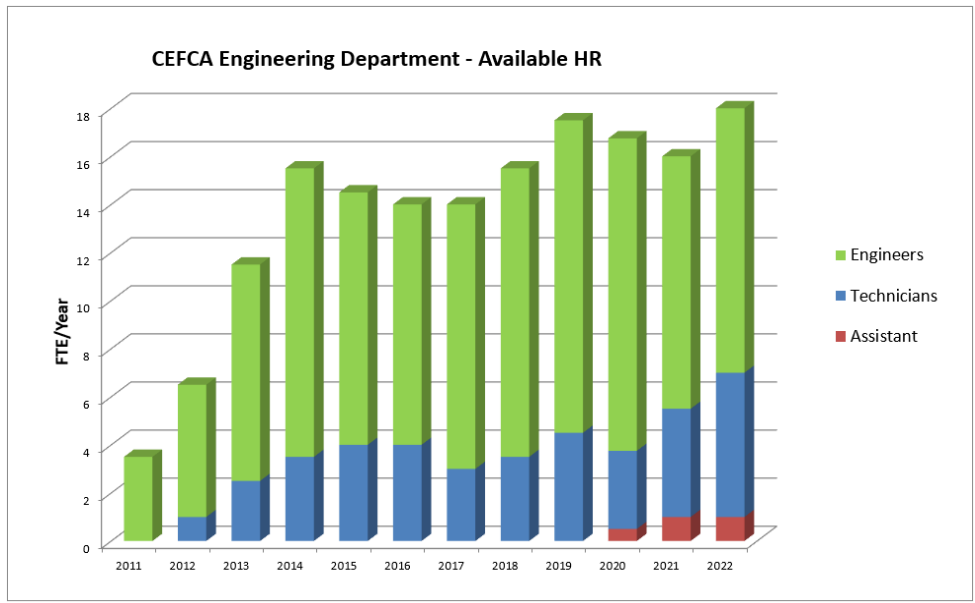

**Figure 11**

Figure 11 shows historical HR of engineering department (FTE/year)

Take into account that engineering staff has been optimized to perform design, developments, implementations and operate the observatory for working mode service and maintenance, as well as general maintenance works and support implementation tasks for CEFCA Headquarters and Galactica, which is a Centre for Astronomy, Education and Outreach located near the OAJ providing access to professional quality facilities for environmental and culture education, scientific and outreach purposes. All of which, make a high work performance demand of engineering team to satisfy all CEFCA infrastructures needs with this high optimized staff.

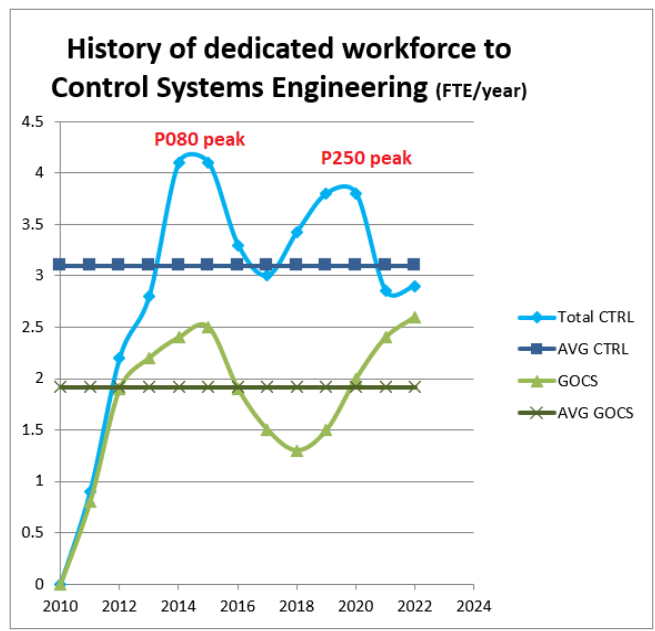

**Figure 12**

Figure 12 shows the dedicated workforce history to control engineering (FTE/year)

The blue curves showed in figure 12 represents the total FTE per year dedicated to control engineering and its average. The green curves represent partial workforce as FTE per year dedicated exclusively to GOCS design, development, implementation, operations (support, service, technical, maintenance) and its average as well. This manpower takes into account partial dedication of these HR profiles depending on project state: control architecture engineer, control hardware engineers and control software engineers.

Apart from GOCS, CEFCA control group has developed a large number of control systems for the observatory. This fact outstands itself if you take into account that the difference between the blue and the green curve is directly the dedication to systems (exclusively local control systems developed by CEFCA) this dedication does not include the part of GOCS to integrate the system.

We achieved, at the end of 2014, the minimum functionality of GOCS to operate the observatory with 2 support process channels (PINS,PGEN) and one scientific channel, the P080. The total workforce to reach this milestone was 10 FTE. We would like to pinpoint that the classical functionality of "OCS et al" approach was reached as part of this milestone, but also extra functionality based on GOCS infrastructure deployed at the observatory and both support process channels.

Assuming that the complete development of GOCS with full planned functionality would be reached in 2024, we can calculate the total FTE that we get as a result, which is a total amount of 28 FTE for completing the GOCS.

**P080 Science process channel (25 integrated subsystems):** in charge of performing the scientific operation of the JAST/T80 telescope related set of systems. As an example, you can see in figure 3 a representative part of these systems.

**P250 Science process channel (37 integrated subsystems):** in charge of performing the scientific operation of the JST/T250 telescope set of systems.

**PGEN General process channel (125 integrated subsystems):** General electronic services such as networks, servers, communications, GOCS observatory control system, UPAD unit for processing and data archiving, alarms, public address system, CCTV security cameras, GPS, access control, lighting, weather station, control rooms, etc.

**PINS Installation process channel (62 integrated subsystems):** Set of coordinated supply installations systems such as water supply, glycol water, water treatment plant, air conditioning, compressed air, LN2 plant, aluminizing vacuum chamber, fire suppression, electrical generators, electrical distribution, electrical consumption, Uninterruptible Power Supply, etc.

## 9. CONCLUSIONS

The CIA[4]-[9] model is applicable to the design of the global control system of any astrophysical observatory, in our case we have implemented it for the Observatorio Astrofísico de Javalambre. The OAJ GOCS[4]-[9] has taken advantage of CIA[4]-[9] rules and recommendations in order to improve uptime, quality and yields; gain real-time visibility to performance parameters and access directly to quality details for observatory operation control, fulfilling scientific and technical standards compliance.

This paper presents a global overview of cost and benefits for developing and implementing a control system engineering project for an astrophysical observatory. We highlight GOCS engineering design based on CIA[4]-[9] model in order to boost operations trying to cover all important parts for showing a global idea of goals, architecture, features and development.

As an overall conclusion, we can assure you that the economic investment of the "GOCS" costs very little compared to the benefits obtained, however, a significant effort must be made in human resources, since dedication to highly specialized tasks is necessary to get the most out of it and higher performance of the entire observatory as a whole.

As the best strategy to follow, we recommend first implementing basic infrastructures and continue gradually implementing the developments based on the human resources available or based on those that the project can assume.

Wanting to implement full functionality in a short time requires a strong investment in human resources or outsourcing the project. However, we do not recommend outsourcing this type of development, since the most important benefit is that organization's own human resources become independent, to optimize GOCS's all process channels taking over the full observatory control to guarantee best performance and offering high quality data for observatory science. It is highly desirable that the knowledge always remains within the organization to get the most out of the tool in such complex scientific-technical processes, because in many cases we usually need to achieve the maximum precision that we can get from coordinated systems as a whole.

A great effort has been made to set up the basic infrastructure of the GOCS, develop the observatory systems and main process channels. Although the GOCS is still passing through a mature period of development, we are now ready to finish the third order control engineering progression, where observatory operation will notice and enjoy the upcoming features and functionalities. The development team is prepared to continue with the following steps in order to finish full management functionality for scientific P250 process channel with JPCam mainly as well as ERP & CMMS & LITE global integration.

Results related to low level infrastructure, local control systems, middleware integration support process channels PGEN, PINS and scientific P080, P250, PMON process channels deployed at OAJ have already demonstrated its great potential. Some extra work is still needed in the final stages of GOCS to achieve the 100% of functionality. However, we are currently enjoying this 92% of GOCS implementation improving, on a daily basis, the efficient OAJ operation as a global concept of fully automated observatory. More effort is required on optimizing the performance of sub-system controls, adjusting control tuning, deploying better configurations for process channels and the experience obtained by the team in the following steps are crucial to reach soon our final goals.

Currently the OAJ GOCS undoubtedly is a first-class development working properly at the Observatorio Astrofísico de Javalambre. This gives us a lot of strength to continue with the next steps in development. Hopefully we can show you in detail in the coming years.

These years of performing scientific operation with GOCS have been exciting because we have been rewarded with successful scientific results published at *The CEFCA Catalogues Portal*[20], with data releases such as J-PLUS EDR, J-PLUS DR1, J-PLUS DR2, J-PAS-PDR201912 and coming soon J-PLUS DR3. These data releases have already produced a series of scientific papers that demonstrate the capabilities of these surveys and the quality of the data.

Should you require further information, do not hesitate to write first author. You are more than welcome to visit us to directly check our GOCS implementation at OAJ.

## ACKNOWLEDGMENTS

Funding for OAJ, UPAD, and CEFCA has been provided by the Governments of Spain and Aragón through the Fondo de Inversiones de Teruel; the Aragonese Government through the Research Groups E96, E103, E16_17R, and E16_20R; the Spanish Ministry of Science, Innovation and Universities (MCIU/AEI/FEDER, UE) with grant PGC2018-097585-B-C21; the Spanish Ministry of Economy and Competitiveness (MINECO/FEDER, UE) under AYA2015-66211-C2-1-P, AYA2015-66211-C2-2, AYA2012-30789, and ICTS-2009-14; and European FEDER funding (FCDD10-4E-867, FCDD13-4E-2685).

The Fundación ARAID of the Government of Aragón and the programs Ramón y Cajal and Juan de la Cierva of the spanish MINECO are acknowledged for their contribution to the project. Funding for the J-PAS project has been provided also by the Brazilian agencies FINEP, FAPESP, FAPERJ and by the National Observatory of Brazil. The contributions from Caja Rural de Teruel, SC and IBERCAJA S.A are also acknowledged.

The first author would like to acknowledge the dedicated work of the entire CEFCA team over these years. The remarkable success of the current state of the observatory is the result of a true team effort. This visible success gives us added energy to continue with the successful operation of the observatory and its data exploitation.

Finally, the authors would like to thank Lluis Moreno from Beckhoff Spain and Yolanda Marco Villanueva, from Teruel, Spain, for their valuable support. Their kind attention has been much appreciated.

# Citation <u>Download Citation</u>

A. Yanes-Díaz, S. Rueda-Teruel, R. Bello, D. Lozano-Pérez, M. Royo-Navarro, T. Civera, M. Domínguez-Martínez, N. Martínez-Olivar, S. Chueca, C. Iñiguez García, A. Marín-Franch, F. Rueda-Teruel, G. López-Alegre, S. Bielsa, J. Muñoz-Maudos, H. Rueda-Asensio, A. Muñoz-Teruel, D. Garcés-Cubel, I. Soriano-Laguía, M. Almarcegui-Gracia, A. J. Cenarro, M. Moles, D. Cristobal-Hornillos, J. Varela, A. Ederoclite, H. Vázquez Ramió, M.C. Díaz-Martín, R. Iglesias-Marzoa, J. Castillo, A, López-Sainz, J. Hernández-Fuertes, D. Muniesa-Gallardo, A. Moreno-Signes, A. Hernán-Caballero, C. López-Sanjuan, A. del Pino, M. Akhlaghi, I. Pintos-Castro, J. Fernández-Ontiveros, F. Hernández-Pérez, S. Pyrzas, R. Infante-Sainz, T. Kuutma, D. Lumbreras-Calle, N. Maicas-Sacristan, J. Lamadrid-Gutierrez, F. López-Martinez, P. Galindo-Guil, E. Lacruz-Calderón, L. Valdivielso-Casas, M. Aguilar-Martín, S. Eskandarlou, A. Domínguez-Fernández, F. Arizo-Borillo, S. Vaquero-Valer, I. Muñoz-Igado, M. Alegre-Sánchez, G. Julián-Caballero de España, A. Romero, and D. Casinos-Cardo "Costs and benefits of automation for astronomical facilities", Proc. SPIE 12186, Observatory Operations: Strategies, Processes, and Systems IX, 121860Y (25 August 2022); https://doi.org/10.1117/12.2626105

# Copyright: